\documentclass[final,1p,times,number]{elsarticle}

\usepackage{amsmath}
\usepackage{amssymb}
\usepackage{amsfonts}
\usepackage{graphicx}
\usepackage{epstopdf}
\usepackage{color}
\usepackage{bm}
\usepackage{multirow}
\usepackage{natbib}
\usepackage{hyperref}

\usepackage{ifpdf}
\ifpdf%
\usepackage{pdflscape}
\else
\usepackage{lscape}
\fi

\journal{Annals of Physics}
\biboptions{sort&compress}

\begin{document}

\begin{frontmatter}
\title{Differential Poisson's ratio of a crystalline two-dimensional membrane}
\author[ISB1,ISB2,IVG1,IVG2]{I.S. Burmistrov\corref{cor}}
\cortext[cor]{Corresponding author. Fax: +7-495-7029317 }
\ead{burmi@itp.ac.ru}
\author[IVG3,IVG1,ISB1]{V. Yu.~Kachorovskii}
\author[IVG1,IVG2,IVG3,ISB1]{I.V. Gornyi}
\author[IVG1,IVG2,ADM1,ISB1]{A.D. Mirlin}
\address[ISB1]{L.D. Landau Institute for Theoretical Physics,
Kosygina street 2, 119334 Moscow, Russia}
\address[ISB2]{\mbox{Laboratory for Condensed Matter Physics, National Research University Higher School of Economics, 101000 Moscow, Russia}}
\address[IVG1]{Institut f\"ur Nanotechnologie,  Karlsruhe Institute of Technology,
76021 Karlsruhe, Germany}
\address[IVG2]{\mbox{Institut f\"ur Theorie der kondensierten Materie,  Karlsruhe Institute of
Technology, 76128 Karlsruhe, Germany}}
\address[IVG3]{A. F.~Ioffe Physico-Technical Institute,
194021 St.~Petersburg, Russia}
\address[ADM1]{Petersburg Nuclear Physics Institute, 188300, St.Petersburg, Russia}

\begin{abstract}
We compute the differential Poisson's ratio of a suspended two-dimensional crystalline membrane embedded into a space of large dimensionality $d \gg 1$. 
We demonstrate that, in the regime of anomalous Hooke's law, the differential Poisson's ratio approaches a universal value determined solely by the spatial dimensionality $d_c$, with a power-law expansion
$\nu = -1/3 + 0.016/d_c + O(1/d_c^2)$, where $d_c=d-2$. 
Thus, the value $-1/3$ predicted in previous literature holds only in the limit $d_c\to \infty$.
\end{abstract}


\end{frontmatter}

\section{Introduction}
\label{sec:intro}

Poisson's ratio is defined as the ratio of a transverse compression to a longitudinal stretching. 
In the classical theory of elasticity, the Poisson's ratio is given by 
$$\nu_\text{cl}=\frac{\lambda}{2\mu+(D-1)\lambda},$$ 
where $\mu$ and $\lambda$ are the Lam\'e coefficients and $D$ is the dimensionality of the elastic body \cite{LandauLifshitz}. 
General conditions of thermodynamic stability restrict the Poisson's ratio to the range between $-1$ and $1/(D-1)$. 
Conventionally, a material contracts in transverse directions when it is stretched in the longitudinal direction, 
such that the Poisson's ratio is positive. However, some exotic, so-called auxetic \cite{evans}, materials     
have a negative Poisson's ratio. Although examples of such materials, e.g., pyrite, have been known for a long time \cite{pyrite}, 
the interest to auxeticity started only at the end of 1980s after the observation of a stretching-induced transverse 
expansion of polyurethane foam \cite{foam}. Nowadays, a negative Poisson's ratio is found in various materials 
and artificially engineered structures (see Ref.~\cite{auxetic} for a review).

An interesting example of auxetic material is a crystalline membrane of dimension $D$ embedded into the space of dimension $d > D$. 
The self-consistent theory of such crystalline membranes \cite{Doussal} predicts the negative Poisson's ratio in the thermodynamic limit. 
This limit is achieved in large membranes, when the membrane size $L$ exceeds the Ginzburg length $L_* \sim \varkappa/\sqrt{T \mu}$, where $\varkappa$ 
is the bending rigidity and $T$ stands for the temperature. A crystalline membrane hosts $d_c=d-D$ soft out-of-plane modes, the so-called flexural phonons, 
which are characterized by strong anharmonicity mediated by the coupling to conventional in-plane phonons \cite{Nelson}. 
As a consequence of such anharmonicity, the elastic moduli show a nontrivial power-law scaling with the system size, temperature, and tension. 
The scaling of all elastic moduli, $\lambda, \mu, \varkappa$ is controlled by the universal exponent $\eta$ which depends only on $d_c$.
The critical exponent $\eta$ was determined within several approximate analytical schemes\cite{david1,david2,Aronovitz89,Doussal,eta1},  
none of which being controllable in the physical case $D=2$ and $d=3$. Numerical simulations for the latter case yielded $\eta=0.60 \pm 0.10$ \cite{Gompper91}, $\eta=0.72 \pm 0.04$ \cite{Bowick96}, and $\eta=0.85$   \cite{Los-PRB-2009}. 
It is because of the nontrivial scaling of the elastic moduli that the linear Hooke's law fails in the regime of small tension \cite{buck,lower-cr-D2,david2,nicholl15,nelson15,katsnelson16,my-hooke}.

%Existence of the anomalous, non-linear Hooke's law results in difference between absolute and differential Poisson's ratio, as is shown by the present authors, together with Katsnelson and Los, in a parallel paper \cite{our-PR-paper}. 

Le Doussal and Radzihovsky \cite{Doussal} found a negative Poisson's ratio  of a two-dimensional crystalline membrane within the self-consistent screening approximation. More specifically, they obtained an entirely universal value $\nu=-1/3$ independent of the spatial dimensionality $d_c$. 
In Ref. \cite{nelson15}, this result of the self-consistent membrane theory was reproduced by Kosmrlj and Nelson by means
of a renormalization-group analysis for a relatively large membrane size $L\gg L_*$ and not too strong tension, $\sigma \ll \sigma_* = \varkappa L_*^{-2}$. 
On the other hand, as shown by the present authors together with Katsnelson and Los in a parallel paper  \cite{our-PR-paper}, the Poisson's ratio is strongly dependent of boundary conditions in the range of lowest tensions (linear-response regime), $\sigma \gtrsim \sigma_L = \varkappa L^{\eta-2}L_*^{-\eta}$.  An independence on boundary conditions is reached only at stronger tensions, $\sigma \gg \sigma_L$. However, also in this regime, one should exert a care when defining the Poisson's ratio. Specifically, emergence of the anomalous, non-linear Hooke's law results in an essential  difference between the absolute and differential Poisson's ratio, as shown in Ref.~\cite{our-PR-paper}.

In this paper, we consider the non-linear regime $\sigma_L \ll \sigma \ll \sigma_*$ and focus on the differential Poisson's ratio. In order to define the differential Poisson's ratio $\nu$, one needs to consider the response to an infinitesimally small anisotropic tension: 
$\sigma_{\parallel}= \sigma+\delta\sigma$ and $\sigma_{\perp}= \sigma$. Then, the ratio of the infinitesimally small change in transverse, $\delta \varepsilon_\perp$, and longitudinal, $\delta \varepsilon_\parallel$, stretching determines the differential Poisson's ratio
\begin{equation}
\nu = - \frac{\delta \varepsilon_\perp}{\delta \varepsilon_\parallel}.
\label{eq:main:def}
\end{equation}
We demonstrate that in the regime $\sigma_L \ll \sigma \ll \sigma_*$ the differential Poisson's ratio indeed acquires a universal value. However, contrary to the result of the self-consistent membrane theory, this universal value depends on the dimensionality $d_c$ of embedded space. We perform calculations which are controlled by the small parameter $1/d_c\ll 1$ and find that the differential Poisson's ratio of the two-dimensional crystalline membrane is given by
\begin{equation}
\nu =  -\frac13+ \frac{0.016}{d_c} + O\bigl (d_c^{-2}\bigr ) \ ,
\qquad \sigma_L \ll \sigma \ll \sigma_* .
\label{eq:nu:main}
\end{equation}
Thus, the differential Poisson's ratio at $\sigma_L \ll \sigma \ll \sigma_*$ is universal (in the sense of independence on material parameters)  but represents a nontrivial function of $d_c$.

The paper is organized as follows. In Sec. \ref{s1} we present the general formalism for the computation of the differential Poisson's ratio of a two-dimensional crystalline membrane. The details of evaluation of the differential Poisson's ratio to the first order in $1/d_c$ are presented in Sec. \ref{s2}. We end the paper with a summary of results, Sec. \ref{s3}. Technical details are given in Appendices.

%%%%%%%%%%%%%%%%%%

\section{Formalism\label{s1}}

We start with the partition function of a two-dimensional crystalline membrane written in terms of the functional integral over in-plane, $\bm{u} = \{u_x,u_y\}$, and out-of-plane, $\bm{h}=\{h_1,\dots ,h_{d_c}\}$ phonons (see Refs. \cite{my-crump,my-hooke,our-Thermal}):
\begin{equation}
Z = \int \mathcal{D}[\bm{u},\bm{h}]\, \exp (-S) .
\label{eq:PFZ}
\end{equation}
Here the action in the imaginary time is given by ($\beta=1/T$)
\begin{gather}
S = \int\limits_0^\beta d\tau \int d^2 \bm{x} \Biggl \{\Bigl [ \frac{\mu}{4}\delta_{\alpha\beta}+\frac{\lambda}{8}\Bigr ] \Bigl [  \left (\xi_\alpha^2-1+\overline{K}_\alpha\right )\left (\xi_\beta^2-1+\overline{K}_\beta\right ) - \overline{K}_\alpha \overline{K}_\beta  \Bigr ]
 +\frac{\rho}{2} \Bigl [ (\partial_\tau \bm{u})^2 +(\partial_\tau \bm{h})^2\Bigr ] \notag \\
 +
\frac{\varkappa}{2}\Bigl [ (\Delta \bm{h})^2+(\Delta \bm{u})^2\Bigr ] + \mu {u}_{\alpha\beta}  {u}_{\beta\alpha}
+\frac{\lambda}{2} {u}_{\alpha\alpha} {u}_{\beta
\beta} \Biggr \} ,
\label{eq:action:i}
\end{gather}
where
\begin{equation}
{u}_{\alpha\beta} = \frac{1}{2} \Bigl ( \xi_\beta \partial_\alpha u_\beta +\xi_\alpha \partial_\beta u_\alpha + \partial_\alpha \bm{u}  \partial_\beta \bm{u} + \partial_\alpha \bm{h} \partial_\beta \bm{h} \Bigr ) ,
\end{equation}
and
\begin{equation}
\overline{K}_\alpha =\frac{1}{\beta L^2} \int\limits_0^\beta d\tau \int d^2 \bm{x}\, K_\alpha, \qquad
K_\alpha = \partial_\alpha \bm{u}  \partial_\alpha \bm{u} + \partial_\alpha \bm{h} \partial_\alpha \bm{h}  .
\end{equation}

The free energy per unit area, $f = -T L^{-2}\ln Z$, is a function of the stretching factors $\xi_x$ and $\xi_y$, i.e. $f\equiv f(\xi_x,\xi_y)$. With the function $f(\xi_x,\xi_y)$, the diagonal components of the tension tensor can be found as
\begin{equation}
\sigma_x = \frac{1}{\xi_x} \frac{\partial f}{\partial \xi_x}, \qquad
\sigma_y = \frac{1}{\xi_y} \frac{\partial f}{\partial \xi_y} .
\label{eq:EOS}
\end{equation}
We emphasize that Eq. \eqref{eq:EOS} determines the dependence of the tension tensor $\{\sigma_x, \sigma_y\}$ on the stretching tensor $\{\xi_x,\xi_y\}$, i.e., Eq. \eqref{eq:EOS} is the equation of state.

In order to find the differential Poisson's ratio $\nu$, we consider the case of slightly anisotropic stretching factors, 
$\xi_\alpha = \xi+\delta\varepsilon_\alpha$, and adjust the ratio $\nu= - \delta\varepsilon_y/\delta\varepsilon_x$ in such a way 
that the components of the tension tensor, $\sigma_x= \sigma+\delta\sigma$ and $\sigma_y= \sigma$, differ only by an infinitesimal addition $\delta\sigma$ in $\sigma_x$.
\color{black}
Then, we find
\begin{equation}
\nu  = \frac{\displaystyle\left (\frac{\partial \sigma_y}{\displaystyle\partial\xi_x}\right )_{\xi_y}}{\displaystyle\left (\frac{\partial \sigma_y}{\partial\xi_y}\right )_{\xi_x}}= \frac{\displaystyle\frac{\partial^2 f}{\partial\xi_y\partial\xi_x}}{\displaystyle\frac{\partial^2 f}{\partial\xi_y^2}-\sigma}.
\label{eq:gen}
\end{equation}
Here the derivatives are taken at $\xi_x=\xi_y=\xi$.

We note that instead of independent variables $\xi_x$ and $\xi_y$, one can choose as independent variables the components of the tension tensor, $\sigma_x$ and $\sigma_y$. Equation \eqref{eq:gen} can be then rewritten in an alternative form:
\begin{equation}
\nu
= - \frac{\displaystyle \left (\frac{\partial\xi_y}{\partial \sigma_x}\right )_{\sigma_y}}{\displaystyle\left (\frac{\partial\xi_x}{\partial \sigma_x}\right )_{\sigma_y}}= - \frac{\displaystyle\frac{\partial^2 g}{\displaystyle\partial \sigma_x\partial \sigma_y}}{\displaystyle\frac{\partial^2 g}{\partial \sigma_x^2}} ,
\label{eq:PR:alt}
\end{equation}
where the derivatives are assumed to be calculated for $\sigma_x=\sigma_y=\sigma$. As usual, the free energy $g(\sigma_x,\sigma_y)$ is related to the free energy $f(\xi_x,\xi_y)$ via the Legendre transform:
\begin{equation}
g(\sigma_x,\sigma_y) = f(\xi_x,\xi_y) - \sigma_x (\xi^2_x-1)/2-\sigma_y(\xi_y^2-1)/2 ,
\end{equation}
where $\xi_\alpha$ is expressed in terms of $\sigma_\alpha$ with the help of the equation of state \eqref{eq:EOS}.
We note that the expression \eqref{eq:PR:alt} has been used for the numerical evaluation of the Poisson's ratio in Ref. \cite{Bowick97}
(though with the different form of the free energy).
Although, both formulations \eqref{eq:gen} and \eqref{eq:PR:alt} are completely equivalent, in what follows we will use the formulation in which the stretching factors $\xi_\alpha$ are the independent variables.

Using the exact form \eqref{eq:action:i} of the action, one finds the following expressions for the second derivatives of the partition function $f$:
\begin{gather}
\frac{\partial^2 f}{\partial\xi_y\partial\xi_x}\Biggl |_{\xi_x=\xi_y=\xi} =
\xi^2 \lambda - \xi^2 \int d\tau^\prime d\bm{x}^\prime \langle \langle L_y(\bm{x},\tau) \cdot L_x(\bm{x}^\prime,\tau^\prime) \rangle \rangle ,
\label{eq:eq1} \\
\frac{\partial^2 f}{\partial\xi^2_y}\Biggl |_{\xi_x=\xi_y=\xi} = \sigma +
\xi^2 (2\mu+\lambda) + (2\mu+\lambda) \bigl\langle (\partial_y u_y)^2\bigr\rangle +\mu \bigl\langle (\partial_x u_y)^2\bigr\rangle
- \frac{2\mu+\lambda}{\xi} \langle  {u}_{yy} \partial_y u_y \rangle
\notag \\
 - \frac{2\mu}{\xi} \langle  {u}_{xy} \partial_x u_y \rangle
-\xi^2 \int d\tau^\prime d\bm{x}^\prime \langle \langle L_y(\bm{x},\tau) \cdot L_y(\bm{x}^\prime,\tau^\prime) \rangle \rangle .
\label{eq:eq2}
\end{gather}
Here the average $\langle \dots \rangle$ is defined with respect to the action \eqref{eq:action:i}, $\langle\langle A \cdot B \rangle \rangle = \langle A B\rangle - \langle A\rangle \langle B\rangle$ and
\begin{gather}
L_x = \frac{2\mu+\lambda}{2} K_x + \frac{\lambda}{2} K_y + \frac{2\mu+\lambda}{2\xi}  {u}_{xx} \partial_x u_x
+ \frac{2\mu}{\xi}   {u}_{yx} \partial_y u_x +\frac{\lambda}{\xi}  {u}_{yy} \partial_x u_x , \\
L_y = \frac{2\mu+\lambda}{2} K_y + \frac{\lambda}{2} K_x + \frac{2\mu+\lambda}{2\xi}  {u}_{yy} \partial_y u_y
+ \frac{2\mu}{\xi}  {u}_{xy} \partial_x u_y +\frac{\lambda}{\xi}  {u}_{xx} \partial_y u_y .
\end{gather}
Equations \eqref{eq:gen}, \eqref{eq:eq1}, and \eqref{eq:eq2} express the Poisson's ratio in terms of correlation functions of elastic deformations. The actual computation of these correlation functions of the in-plane and flexural phonons is complicated due to interaction between these phonon modes.

Below we limit the analysis to the case of high temperature, $T\gg \varkappa^2/(\mu L^2)$ in which one can consider the phonons to be quasistatic. In this regime, one can also neglect the term $\partial_\alpha \bm{u} \partial_\beta \bm{u}$ in comparison with  $\partial_\alpha \bm{h} \partial_\beta \bm{h}$ in the expressions for $\tilde{u}_{\alpha\beta}$ and $K_\alpha$. Then we can simplify Eqs. \eqref{eq:eq1} and \eqref{eq:eq2}. Indeed, by making the following change of variables: $u_\alpha \to \xi_\alpha u_\alpha$, we can recast the partition function \eqref{eq:PFZ} as:
\begin{equation}
Z = \int \mathcal{D}[\bm{u},\bm{h}]\, \exp (-\tilde{E}/T) ,
\end{equation}
where
\begin{gather}
\tilde{E} =  \int d^2 \bm{x} \Biggl \{
\Biggl [ \frac{\mu}{4} \delta_{\alpha\beta} +\frac{\lambda}{8}\Biggr ]
\Biggl [  \left (\xi_\alpha^2-1+\overline{\tilde K}_\alpha\right )\left (\xi_\beta^2-1+\overline{\tilde K}_\beta\right ) - \overline{\tilde K}_\alpha \overline{\tilde K}_\beta  \Biggr ]
 +
\frac{\varkappa}{2} (\Delta \bm{h})^2  \notag \\
+ \mu \tilde{u}_{\alpha\beta}  \tilde{u}_{\beta\alpha}
+\frac{\lambda}{2} \tilde{u}_{\alpha\alpha} \tilde{u}_{\beta
\beta} \Biggr \} .
\label{eq:action:ii}
\end{gather}
Here we have introduced the following notations:
\begin{equation}
\tilde{u}_{\alpha\beta} = \frac{1}{2} \Bigl ( \partial_\alpha u_\beta + \partial_\beta u_\alpha +  \partial_\alpha \bm{h} \partial_\beta \bm{h} \Bigr ), \qquad  \overline{\tilde K_\alpha} = \frac{1}{L^2}\int d^2\bm{x} \tilde{K}_\alpha, \qquad \tilde{K}_\alpha = \partial_\alpha \bm{h} \partial_\alpha \bm{h}  .
\end{equation}

Since the action \eqref{eq:action:ii} becomes quadratic in the in-plane phonons, we can integrate them out and express the partition function as an integral over static flexural phonons,
\begin{equation}
Z = \int \mathcal{D}[\bm{h}]\, \exp (-E/T) ,
\end{equation}
where the energy $E$ for a given configuration of the flexural phonon field $h(\bm{x})$ is given by \cite{my-hooke}
\begin{align}
E = &  \int d^2 \bm{x}\ \Biggl [ \frac{\mu}{4}\delta_{\alpha\beta}+\frac{\lambda}{8}\Biggr ] \Biggl [  \left (\xi_\alpha^2-1+\overline{\tilde K_\alpha}\right )\left (\xi_\beta^2-1+\overline{\tilde K_\beta}\right ) \Biggr ]
  + \frac{\mu}{2 L^2}\left ( \int d^2 \bm{x}  \
  \partial_x \bm{h} \partial_y \bm{h}\right )^2
   \notag \\
& +\frac{\varkappa}{2} \int d^2 \bm{x}\  (\Delta \bm{h})^2+ \frac{2\mu(\mu+\lambda)}{4(2\mu+\lambda)} \int^\prime \frac{d^2 \bm{k} d^2\bm{k^\prime} d^2 {\bm q}}{(2\pi)^6}
  \frac{[\bm{k}\times \bm{q}]^2}{q^2}\frac{[\bm{k^\prime}\times \bm{q}]^2}{q^2} \bigl (\bm{h}_{\bm{k}+\bm{q}} \bm{h}_{-\bm{k}}\bigr )\bigl (
  \bm{h}_{-\bm{k^\prime}-\bm{q}} \bm{h}_{\bm{k^\prime}}\bigr ) .
\label{eq:action:iv}
\end{align}
The `prime' sign in the last integral means that the interaction with $q=0$ is excluded:
the `zero-mode' term with $q=0$ from the contributions $\tilde{u}_{\alpha\beta}  \tilde{u}_{\beta\alpha}$ and $\tilde{u}_{\alpha\alpha} \tilde{u}_{\beta
\beta}$ to the energy $\tilde{E}$ in Eq.~(\ref{eq:action:ii}) has been combined with the term $\overline{\tilde K}_\alpha \overline{\tilde K}_\beta$, yielding exactly the term with 
$\partial_x \bm{h} \partial_y \bm{h}$ in Eq.~(\ref{eq:action:iv}). 

Since now $\xi_\alpha$ does not enter the interaction part of the free energy which depends on $\bm{u}$, we obtain a much simpler equation of state:
\begin{gather}
\begin{pmatrix}
\sigma_x\\
\sigma_y
\end{pmatrix}
%=\begin{pmatrix}
%\frac{\partial f}{\xi_x\partial \xi_x}
%\\
%\frac{\partial f}{\xi_y\partial \xi_y}
%\end{pmatrix}
= \frac{1}{2} M
\begin{pmatrix}
\xi_x^2-1+\langle \tilde K_x \rangle \\
\xi_y^2-1+\langle \tilde K_y\rangle
\end{pmatrix}, \qquad
M=\begin{pmatrix}
2\mu+\lambda & \lambda \\
\lambda & 2\mu+\lambda
\end{pmatrix} .
%\Lambda= (1/2) M
\label{eq:112}
\end{gather}
Here the average $\langle\dots \rangle$ is with respect to
the energy \eqref{eq:action:iv}. The second derivatives of the free energy with respect to 
the stretching factors become
\begin{gather}
\frac{\partial^2 f}{\partial\xi_y\partial\xi_x}\Biggl |_{\xi_x=\xi_y=\xi} =
\xi^2 \lambda - \xi^2 \beta \int  d\bm{x}^\prime \langle \langle \tilde{L}_y(\bm{x}) \cdot \tilde{L}_x(\bm{x}^\prime) \rangle \rangle ,
\label{eq:eq1-1} \\
\frac{\partial^2 f}{\partial\xi^2_y}\Biggl |_{\xi_x=\xi_y=\xi} = \sigma +
\xi^2 (2\mu+\lambda)
 -\xi^2\beta  \int  d\bm{x}^\prime \langle \langle \tilde{L}_y(\bm{x}) \cdot \tilde{L}_y(\bm{x}^\prime) \rangle \rangle ,
\label{eq:eq2-1}
\end{gather}
where $\tilde{L}_\alpha = M_{\alpha\beta} \tilde{K}_\beta/2$.

The energy functional $E$ involves two types of interaction of flexural phonons. The terms
% quadratic in $\overline{\tilde K}_\alpha$
in the first line of Eq. \eqref{eq:action:iv} correspond to the interaction with zero momentum transfer
('zero mode'). In the case of large membrane size, $\sigma\gg \sigma_L$, this interaction can be treated in the random phase approximation. Then, we find
\begin{equation}
\frac{\beta}{2} \int d^2\bm{x^\prime} \langle \langle \tilde K_\alpha(\bm{x})\cdot \tilde K_\beta(\bm{x^\prime}) \rangle \rangle = \left [ \Pi \left ( 1+ \frac{1}{2}M \Pi  \right )^{-1}\right ]_{\alpha \beta} ,
 \end{equation}
where $\Pi$ denotes the polarization operator (at zero momentum) irreducible with respect to the interaction with the zero-momentum transfer:
\begin{equation}
\Pi_{\alpha\beta} = \frac{\beta}{2} \int d^2\bm{x^\prime} \langle \langle \tilde K_\alpha(\bm{x})\cdot \tilde K_\beta(\bm{x^\prime}) \rangle \rangle_{\rm irr} .
\end{equation}
We note that $\Pi_{\alpha\beta}$ has two independent components: $\Pi_{xx}=\Pi_{yy}$ and $\Pi_{xy}=\Pi_{yx}$. Using Eqs. \eqref{eq:eq1-1} and \eqref{eq:eq2-1}, we express the differential Poisson's ratio in terms of the components of $\Pi$:
\begin{equation}
\nu = \frac{\nu_0 -Y_0 \Pi_{xy}/2}{1+Y_0  \Pi_{xx}/2} .
\label{eq:PR:m}
\end{equation}
Here
\begin{equation}
\nu_0=\frac{\lambda}{2\mu+\lambda}, \qquad Y_0 = \frac{4\mu(\mu+\lambda)}{2\mu+\lambda}
\end{equation}
denote the bare values of the Poisson's ratio and Young modulus for the two-dimensional crystalline membrane, respectively.

In order to clarify the meaning of $\Pi_{xx}$ and $\Pi_{xy}$,
it is useful to consider a general form of the polarization operator at finite momentum $\bm{q}$:
\begin{equation}
\hat\Pi_{\alpha\beta,\gamma\delta}(\bm{q}) = \frac{1}{2} \int  d^2\bm{x^\prime}\, e^{-i \bm{q}(\bm{x}-\bm{x^\prime})} \langle \langle \bigl ( \nabla_\alpha \bm{h}(\bm{x}) \nabla_\beta \bm{h}(\bm{x}) \bigr )
\cdot  \bigl (\nabla_\gamma \bm{h}(\bm{x^\prime}) \nabla_\delta \bm{h}(\bm{x^\prime})\bigr )\rangle \rangle .
\end{equation}
Due to the rotation symmetry and the symmetry under permutation of the indices $\alpha$ and $\beta$ (as well as $\gamma$ and $\delta$), the polarization operator at zero momentum is expressed as follows \cite{lower-cr-D2}
\begin{equation}
\hat\Pi_{\alpha\beta,\gamma\delta}(0) = \Pi_{xy} \delta_{\alpha\beta}\delta_{\gamma\delta} +
 \frac{1}{2}\bigl (\Pi_{xx}-\Pi_{xy}\bigr )\bigl (  \delta_{\alpha\gamma}\delta_{\beta\delta}+ \delta_{\alpha\delta}\delta_{\beta\gamma} \bigr ) .
\end{equation}
We emphasize that, in general, there are no reasons for $\hat\Pi_{\alpha\beta,\gamma\delta}$ to be fully symmetric with respect to permutations of all its indices as it is assumed in the self-consistent screening approximation \cite{Doussal,Gazit1}. Therefore, Eq. \eqref{eq:PR:m} yields the most general expression for the differential Poisson's ratio. We also note that Eq. \eqref{eq:PR:m} can be written as
(see \ref{App1})
\begin{equation}
\nu = \frac{\lambda^\prime}{2\mu^\prime+\lambda^\prime} ,
\end{equation}
where $\lambda^\prime$ and $\mu^\prime$ are the screened Lam\'e coefficients:
\begin{equation}
\mu^\prime =\frac{\mu}{1+ (\Pi_{xx}-\Pi_{xy})\mu} , \qquad
B^\prime = \frac{B}{1+(\Pi_{xx}+\Pi_{xy}) B} .
\label{eq:screened:elmod}
\end{equation}
Here we have introduced bare and screened bulk moduli: $B=\mu+\lambda$ and $B^\prime=\mu^\prime+\lambda^\prime$, respectively.

In order to find how $\nu$ depends on parameters of the problem, e.g., on the number of flexural phonon modes $d_c$, one needs to compute $\Pi_{xx}$ and $\Pi_{xy}$. In the next section we remind the reader on the results of the self-consistent screening approximation and then compute corrections in $1/d_c$.

Using the equation of state \eqref{eq:112}, we can express the stretching factors $\xi_\alpha$ via tensions $\sigma_\alpha$. Then, with the help of Eq. \eqref{eq:PR:alt}, we find the following representation for the differential Poisson's ratio
\begin{equation}
\nu = \frac{\displaystyle\nu_0+ \frac{Y_0}{2} \left ( \frac{\partial \langle \tilde K_y\rangle}{\partial \sigma_x}\right )_{\sigma_y}}{\displaystyle 1-\frac{Y_0}{2} \left ( \frac{\partial \langle \tilde K_x\rangle}{\partial \sigma_x}\right )_{\sigma_y} } .
\label{eq:nu:25}
\end{equation}
Here $\langle \tilde K_\alpha\rangle$ is expressed in terms of $\sigma_x$ and $\sigma_y$.  After taking derivatives in Eq. \eqref{eq:nu:25}, one sets $\sigma_x=\sigma_y=\sigma$. Below we demonstrate 
how the two representations of the differential Possion ratio, \eqref{eq:PR:m} and \eqref{eq:nu:25}, are related.

The irreducible polarization operator at the zero momentum can be exactly expressed via the full triangular vertex $\Gamma_\beta(\bm{k},\bm{k})$:
\begin{gather}
 \Pi_{\alpha\beta}= \int \frac{d^2 \bm{k}}{(2\pi)^2} k_\alpha^2 \mathcal{G}_k^2 \Gamma_\beta(\bm{k},\bm{k}) .
\end{gather}
Here 
$$\mathcal{G}_k= \frac{T}{\varkappa k^4+
%(\mu+\lambda)(\xi^2-1) k^2 
(\xi^2_\alpha-1)M_{\alpha\beta} k_\beta^2/2- \Sigma_k}$$ 
denotes the propagator for the flexural phonons, with $\Sigma_k$ being the exact self-energy. 
The bare value of the triangular vertex $\Gamma_\beta(\bm{k},\bm{k})$ is equal to $k_\beta^2/T$.
The full triangle vertex satisfies the following identity:
\begin{equation}
\Gamma_\beta(\bm{k},\bm{k}) = \frac{\partial \mathcal{G}_k^{-1}}{\partial \sigma_\beta} .
\label{eq:WI}
\end{equation}
As a consequence of this identity, we obtain
\begin{gather}
 \Pi_{\alpha\beta}= - \frac{\partial}{\partial\sigma_\beta}
\int \frac{d^2 \bm{k}}{(2\pi)^2}\, k_\alpha^2 \mathcal{G}_k = - \frac{\partial\langle \tilde{K}_\alpha\rangle}{\partial\sigma_\beta} .
\label{eq:Pi:GGG}
\end{gather}
Therefore, expressions \eqref{eq:PR:m} and \eqref{eq:nu:25} are identical.

%%%%%%%%%%%%%%%%%%

\section{Evaluation of the differential Poisson's ratio\label{s2}}

\subsection{Self-consistent screening approximation}

The interaction between flexural phonons with finite momentum transfer results in renormalization of the bending rigidity at $k\ll q_*\equiv L_*^{-1}$ \cite{Aronovitz89,lower-cr-D2,Doussal},
\begin{equation}
\varkappa \to \varkappa(k) = \varkappa (q_*/k)^\eta f(k/q_\sigma) .
\label{eq:kappa:ren}
\end{equation}
Here, $q_\sigma = q_* (\sigma/\sigma_*)^{1/(2-\eta)}$ and the function $f(x)$ has the following asymptotic behavior:
\begin{equation}
f(x) =\begin{cases}
1 ,\quad & x\gg 1 ,\\
x^{-\eta}, \quad & x\ll 1 .
\end{cases}
\end{equation}
The simplest approach for computing the irreducible polarization operator is to neglect the vertex corrections. As we shall see below, this can be justified for $d_c\gg 1$. Then, we find
\begin{gather}
 \Pi^{(0)}_{\alpha\beta}= d_c \int \frac{d^2 \bm{k}}{(2\pi)^2 T} \,k_\alpha^2 k_\beta^2\, \mathcal{G}_k^2  .
 \label{eq:Pi:SCSA}
\end{gather}
Independently of the form of the exact propagator $\mathcal{G}_k$, we find the irreducible polarization operator as
\begin{equation}
 \Pi^{(0)}_{\alpha\beta} = d_c \gamma  \langle n_\alpha^2 n_\beta^2\rangle_{\bm{n}}, \qquad
 \gamma = \int \frac{d^2 \bm{k}}{(2\pi)^2 T} \,k^4\, \mathcal{G}_k^2 .
\end{equation}
Here $\bm{n}$ stands for the two-dimensional unit vector and $\langle \dots \rangle_{\bm{n}}$ denotes the averaging over directions of $\bm{n}$. Thus, neglecting the vertex corrections yields the following relation:
\begin{equation}
\Pi^{(0)}_{xx} = 3 \Pi^{(0)}_{xy} .
\label{eq:rel:appr}
\end{equation}
Relation \eqref{eq:rel:appr} implies that $\hat\Pi_{\alpha\beta,\gamma\delta}$ is fully symmetric with respect to permutation of indices. This assumption is used in the self-consistent screening approximation.

Motivated by the renormalization of bending rigidity \eqref{eq:kappa:ren} and the Ward identity (see  \ref{App2}), we use the following ansatz for the exact propagator:
\begin{equation}
{\mathcal{G}_k} = \frac{T}{\varkappa(k) k^4 +\sigma k^2}.
\label{eq:GG}
\end{equation}
The integral over $k$ in Eq. \eqref{eq:Pi:SCSA} is then dominated by $k\sim q_\sigma$ and we obtain
\begin{equation}
 \gamma \sim \frac{T}{\varkappa \sigma} \left (\frac{\sigma_*}{\sigma} \right )^{\eta/(2-\eta)}.
 \end{equation}
Then, from Eq. \eqref{eq:PR:m} we find at $\sigma\ll \sigma_*$ that the differential Poisson's ratio  becomes
\begin{equation}
\nu \approx - \frac{\Pi^{(0)}_{xy}}{\Pi^{(0)}_{xx}} = -\frac{1}{3} .
\label{eq:dPR:univ}
\end{equation}
It is exactly  the result that was obtained within the self-consistent screening approximation \cite{Doussal}.

\subsection{Vertex corrections to the polarization operator}

Corrections to the result \eqref{eq:dPR:univ} stem from the violation of the relation \eqref{eq:rel:appr}. In order to refine the differential Poisson's ratio, we expand the right-hand side of Eq. \eqref{eq:PR:m} in the difference
$3\Pi_{xy}-\Pi_{xx}$:
\begin{equation}
\nu \approx -\frac{1}{3} + \frac{3\Pi_{xy}-\Pi_{xx}}{9 \Pi^{(0)}_{xy}}
\label{eq:nu-with-corr}
\end{equation}
As we shall see below, the correction to the value $-1/3$ will be of the order of $1/d_c$.

\begin{figure}
\centerline{(a) \includegraphics[width=0.3\textwidth]{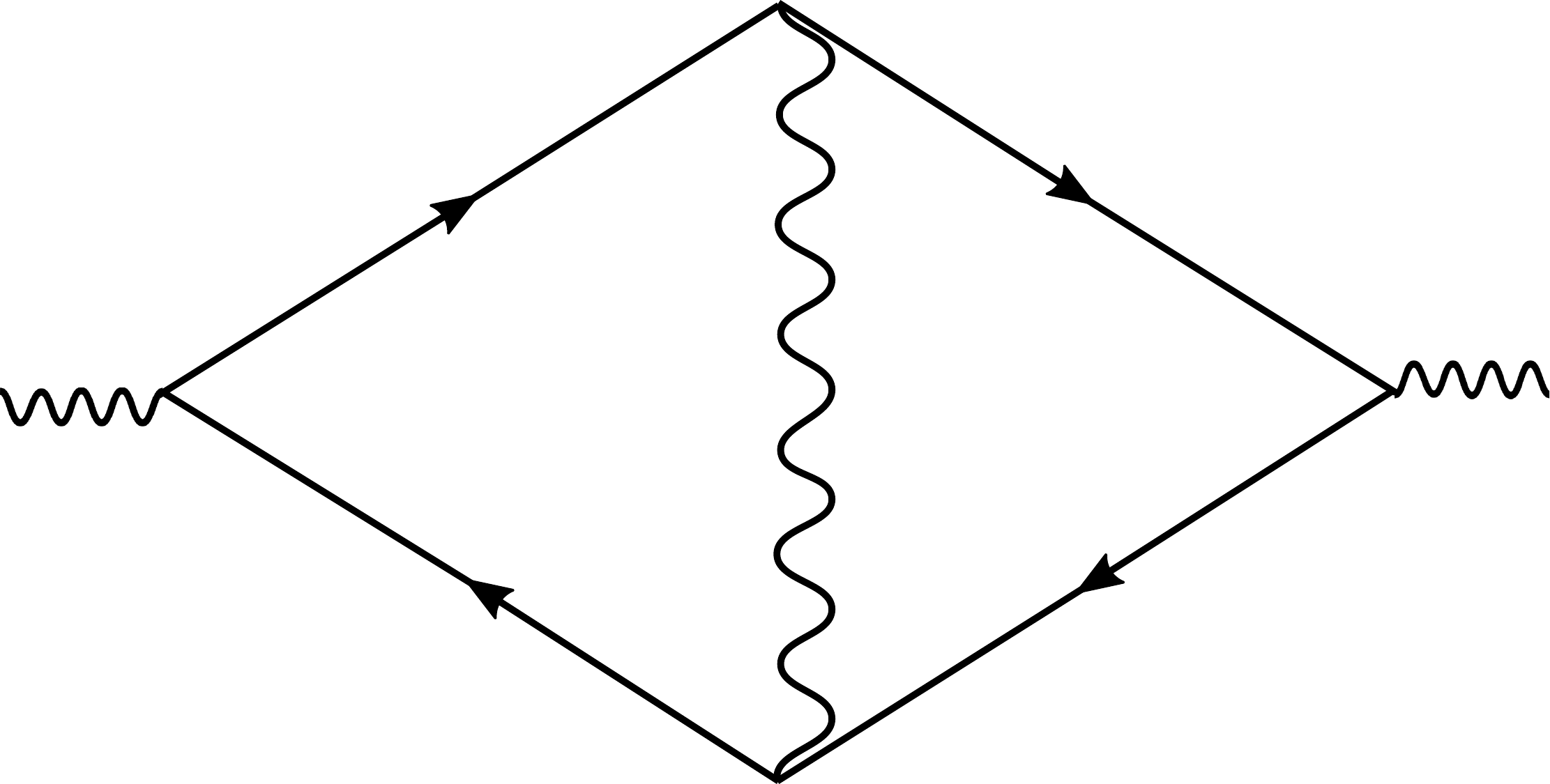}
(b) \includegraphics[width=0.3\textwidth]{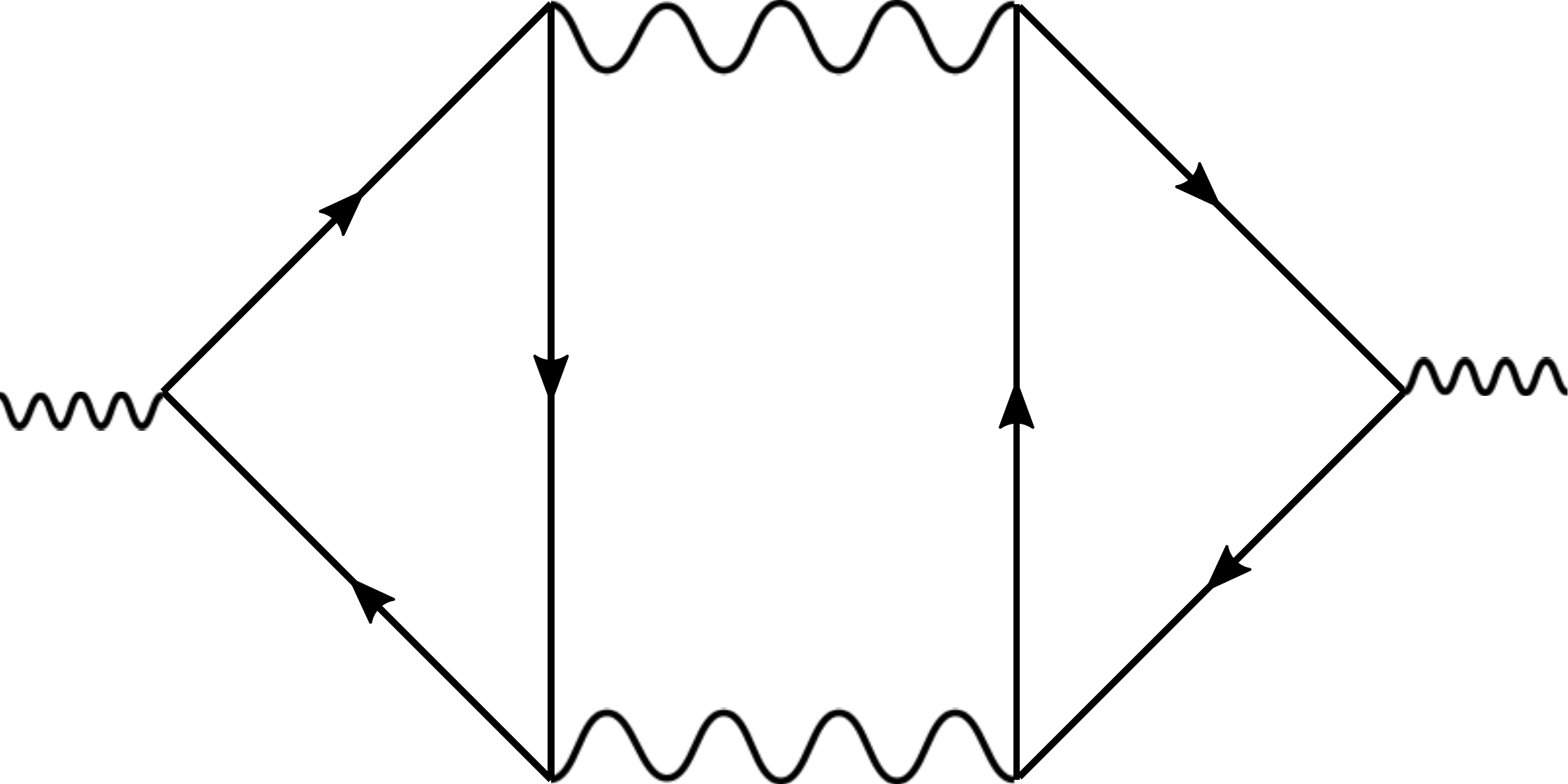} (c) \includegraphics[width=0.3\textwidth]{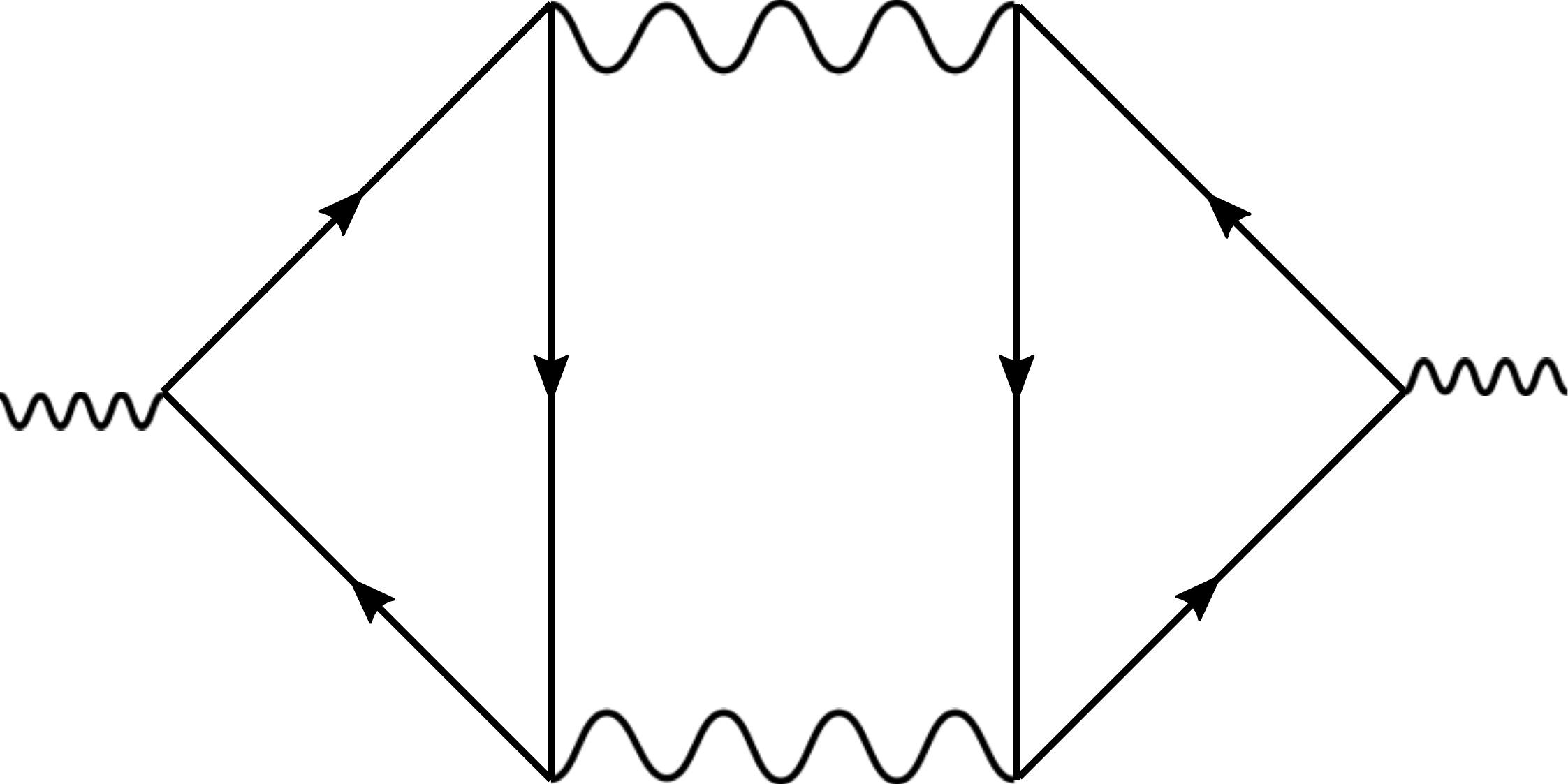}}
\caption{The diagrams of the first order in $1/d_c$ for the polarization operators $\Pi_{xx}$ and $\Pi_{xy}$ at zero momentum transfer.
The solid line denotes the propagator $\mathcal{G}_k$. The wavy line depicts the screened interaction between flexural phonons, which is equal to $1/[3\Pi_q^{(0)}]$ in the universal regime, $q<q_*$.}
\label{fig1}
\end{figure}

There are three diagrams with non-trivial vertex corrections (see Fig. \ref{fig1}) that contribute 
to $\Pi_{\alpha\beta}$ at order $d_c^0$. They yield the following corrections:
\begin{gather}
\Pi^{\rm (a)}_{\alpha\beta} =
- 2 d_c  \int \frac{d^2 \bm{k}d^2 \bm{q}}{(2\pi)^4 T^2} \mathcal{G}_k^2\mathcal{G}_{\bm{k-q}}^2 \frac{[\bm{k}\times \bm{q}]^4}{q^4} N^\prime_q
%\frac{1}{3\Pi^{(0)}_q}
k_\alpha^2 (k_\beta-q_\beta)^{2} ,
\end{gather}
and
\begin{gather}
\Pi^{\rm (b+c)}_{\alpha\beta} =
4 d_c^2 \int \frac{d^2 \bm{k}d^2 \bm{k^\prime}d^2 \bm{q}}{(2\pi)^6 T^3} \mathcal{G}_k^2\mathcal{G}_{\bm{k-q}}\mathcal{G}_{k^\prime}^2\mathcal{G}_{\bm{k^\prime-q}} \frac{[\bm{k}\times \bm{q}]^4}{q^4} \frac{[\bm{k^\prime}\times \bm{q}]^4}{q^4} N_q^{\prime 2}
%\left (\frac{1}{3\Pi^{(0)}_q}\right )^2
k_\alpha^2 k^{\prime 2}_\beta .
\end{gather}
Here $N^\prime_q$ denotes the screened interaction between flexural phonons (see \ref{App1}),
\begin{equation}
N^\prime_q = \frac{Y_0/2}{1+ 3 Y_0 \Pi^{(0)}_q/2} ,
\end{equation}
where $\Pi_q^{(0)}$ stands for the polarization operator at finite momentum calculated without vertex correction. We note that $\Pi^{(0)}_{xx}=3\Pi^{(0)}_{xy} = 3\Pi^{(0)}_{q=0}$.
The polarization operator $\Pi_q^{(0)}$ is given by following expression:
\begin{equation}
\Pi^{(0)}_q = \frac{d_c}{3} \int \frac{d^2 \bm{k}}{(2\pi)^2 T}  \frac{[\bm{k}\times \bm{q}]^4}{q^4} \mathcal{G}_k\mathcal{G}_{\bm{k-q}} .
\label{eq:Pol:Oper}
\end{equation}

Since we are interested in the regime $q\ll q_*$, we can approximate $N^\prime_q$ by $1/[3\Pi^{(0)}_q]$. Then, combining both contributions together, we find
\begin{gather}
\frac{3\Pi_{xy}-\Pi_{xx}}{9 \Pi^{(0)}_{xy}}
=  - \frac{2 d_c}{27\Pi^{(0)}_{xy}}
\int \frac{d^2 \bm{k}d^2 \bm{k^\prime}}{(2\pi)^4 T^2} \mathcal{G}_k^2\mathcal{G}_{k^\prime}^2
\Biggl \{\frac{[\bm{k}\times \bm{k^\prime}]^4}{|\bm{k}- \bm{k^\prime}|^4}\frac{1}{\Pi^{(0)}_{\bm{k}- \bm{k^\prime}}}
-\frac{2d_c}{3} \int \frac{d^2 \bm{q}}{(2\pi)^2 T}\mathcal{G}_{\bm{k-q}}\mathcal{G}_{\bm{k^\prime-q}}\notag \\
\times
\frac{[\bm{k}\times \bm{q}]^4}{q^4} \frac{[\bm{k^\prime}\times \bm{q}]^4}{q^4} \left (\frac{1}{\Pi^{(0)}_q}\right )^2
\Biggr \} \Bigl [ 3 k_x^2 k_y^{\prime 2} - k_x^2 k_x^{\prime 2}\Bigr ] .
\end{gather}
We note that this expression can be written in a rotationaly invariant way. Indeed, in the first term, the expression under the integral sign depends on the angle $\theta$ between $\bm{k}$ and $\bm{k^\prime}$ only. Averaging over directions of $\bm{k}$, we find
\begin{equation}
\int_0^{2\pi} \frac{d\phi}{2\pi} \cos^2\phi\Bigl [ 3\cos^2(\phi+\theta)-\sin^2(\phi+\theta)\Bigr ]  = \sin^2\theta .
\end{equation}
In the second term, the expression under the integral sign depends on the angles $\theta$ and $\theta^\prime$ between $\bm{k}$ and $\bm{q}$, and between $\bm{k^\prime}$ and $\bm{q}$, respectively. Averaging over directions of $\bm{q}$, we find
\begin{equation}
\int_0^{2\pi} \frac{d\phi}{2\pi} \cos^2(\phi+\theta)\Bigl [ 3\cos^2(\phi+\theta^\prime)-\sin^2(\phi+\theta^\prime)\Bigr ]  = \sin^2(\theta-\theta^\prime) .
\end{equation}
Therefore, we obtain
\begin{gather}
\frac{3\Pi_{xy}-\Pi_{xx}}{9 \Pi^{(0)}_{xy}}
= I^{(a)}+I^{(b+c)} ,
\label{eq:gen:corr}
\end{gather}
where
\begin{equation}
\begin{split}
I^{(a)} & =  - \frac{2 d_c}{27\Pi^{(0)}_{xy}}
\int \frac{d^2 \bm{k}d^2 \bm{k^\prime}}{(2\pi)^4 T^2}  \frac{[\bm{k}\times \bm{k^\prime}]^6}{|\bm{k}- \bm{k^\prime}|^4}\frac{\mathcal{G}_k^2\mathcal{G}_{k^\prime}^2}{\Pi^{(0)}_{\bm{k}- \bm{k^\prime}}} ,
%
%\notag
\\
I^{(b+c)}& = \frac{4 d^2_c}{81\Pi^{(0)}_{xy}}
\int \frac{d^2 \bm{k}d^2 \bm{k^\prime}d^2 \bm{q}}{(2\pi)^6 T^3} [\bm{k}\times \bm{k^\prime}]^2\frac{[\bm{k}\times \bm{q}]^4}{q^4 \Pi^{(0)}_q} \frac{[\bm{k^\prime}\times \bm{q}]^4}{q^4 \Pi^{(0)}_q}
\mathcal{G}_{\bm{k-q}} \mathcal{G}_{\bm{k^\prime-q}} \mathcal{G}_k^2\mathcal{G}_{k^\prime}^2
   .
   \end{split}
\label{eq:diff:eq1}
\end{equation}

%%%%%%%%%%%%%%%%%%%%%%%%
%%%%%%%%%%%%%%%%%%%%%%%%

\subsection{Correction to the self-energy}

The results \eqref{eq:gen:corr} and \eqref{eq:diff:eq1} can be derived in a different way using the relation \eqref{eq:WI} between the triangular vertex at zero momentum and the inverse Green's function. In view of Eq. \eqref{eq:Pi:GGG}, in order to find the differential Poisson's ratio one needs to compute the change of the Green's function upon applying an infinitesimally small tension $\delta \sigma$ along the $x$ direction.

In the presence of $\delta \sigma$, the Green's function can be written in terms of the self-energy $\Sigma_{\bm{k}}$:
\begin{equation}
\mathcal{G}_{\bm{k}}=\frac{T}{\varkappa k^4  + \sigma k^2 -
\Sigma_{\bm{k}}} .
\end{equation}
We mention that the ansatz \eqref{eq:GG} used above for $\delta\sigma=0$ corresponds to $\Sigma_{\bm{k}} = [\varkappa-\varkappa(k)]k^4$. We also note that the trivial term $\delta\sigma k_x^2$ is included into $\Sigma_{\bm{k}}$ for the sake of convenience.

\begin{figure}[t!]
\centerline{\includegraphics[width=0.25\linewidth]{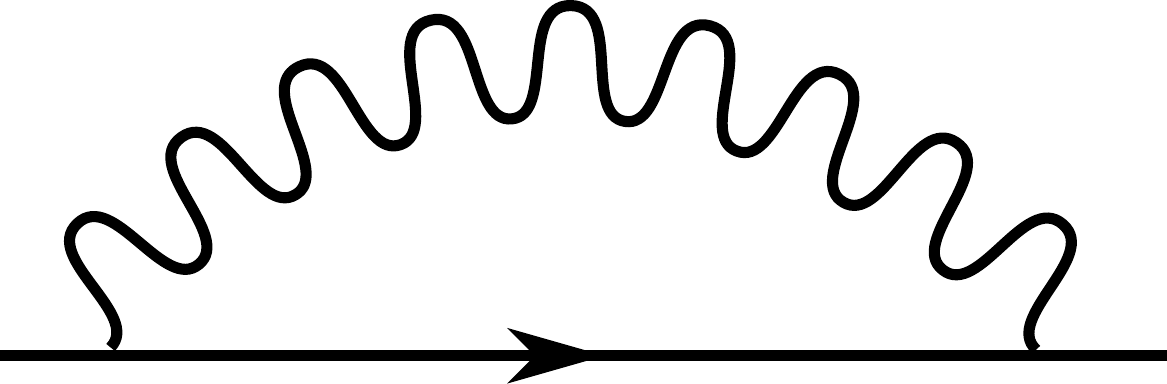}}
\caption{The diagram for the self energy (see text).}
%\vspace*{-0.3cm}
\label{Fig-SE}
\end{figure}

In order to find the change of $\Sigma_{\bm{k}}$ induced by
the infinitesimally small tension $\delta \sigma$, we use the lowest-order diagram for the self-energy (see Fig. \ref{Fig-SE}):
\begin{equation}
\Sigma_{\bm{k}}=\frac{2}{3} \int \frac{d^2\bm{q}}{(2\pi)^2} \frac{[\bm{k}\times\bm{q}]^4}{q^4}
\frac{\mathcal{G}_{\bm{k}-\bm{q}}}{\Pi_q^{(0)}},
\label{Sigma0}
\end{equation}
We note that, as above, the dominant contribution comes from momenta $q\ll q_*$ such that the interaction line is determined by the inverse polarization operator.

As one can see from the diagram in Fig.~\ref{Fig-SE}, the variation of the self-energy in the presence of $\delta \sigma$ arises from the variation of  the Green's function:
\begin{equation}
\delta \mathcal{G}_{\bm{k}-\bm{q}}=
\mathcal{G}_{\bm{k}-\bm{q}}^2 \delta \Sigma_{\bm{k}-\bm{q}}
,
\label{dG}
\end{equation}
as well as from the the change of the polarization operator (see Eq. \eqref{eq:Pol:Oper})
\begin{equation}
\delta \Pi_q^{(0)}= \frac{2 d_c}{3}\int \frac{d^2\bm{k}}{(2\pi)^2}
\frac{[\bm{k}\times\bm{q}]^4}{q^4}
 \delta G_{\bm{k}}  G_{\bm{k}-\bm{q}}=
 \frac{2 d_c}{3}\int \frac{d^2\bm{k}}{(2\pi)^2}
\frac{[\bm{k}\times\bm{q}]^4}{q^4}
 G_{\bm{k}}^2  G_{\bm{k}-\bm{q}}
\delta \Sigma_{\bm{k}}.
\label{dPi}
\end{equation}
Now the correction  $\delta \Sigma_{\bm{k}}$ can be found
from the variation of Eq. \eqref{Sigma0}:
\begin{equation}
\delta \Sigma_{\bm{k}}=-\delta \sigma k_x^2 + \frac{2}{3}
\int \frac{d^2\bm{q}}{(2\pi)^2} \frac{[\bm{k}\times\bm{q}]^4}{q^4} \left [
\frac{\delta \mathcal{G}_{\bm{k}-\bm{q}}}{\Pi_q^{(0)}}
-\frac{\mathcal{G}_{\bm{k}-\bm{q}}\delta \Pi_q^{(0)}}{\bigl[\Pi_q^{(0)}\bigr ]^2}
\right ] .
\label{dSig1}
\end{equation}
Since the right-hand side of this equation is linear in $\delta \Sigma_{\bm k}$, it can be rewritten as
\begin{equation}
(1+\hat \alpha) \delta \Sigma=-\delta \sigma k_x^2 ,
\label{polar}
\end{equation}
where we formally introduce the linear integral operator $\hat \alpha$ as:
  \begin{equation}
 \hat \alpha\, \delta \Sigma_{\bm{k}}=-\frac23 \int\frac{d^2\bm{k^\prime}}{(2\pi)^2}\mathcal{G}^2_{\bm{k^\prime}} \left[ \frac{(\bm{k}\times \bm{k^\prime})^4}{|\bm{k}-\bm{k^\prime}|^4\Pi^{(0)}_{\bm{k}-\bm{k^\prime}}}
 -\frac{2 d_{\rm c}}{3} \int\frac{d^2\bm{q}}{(2\pi)^2}
 \frac{[\bm{k}\times \bm{q}]^4}{q^4 \Pi^{(0)}_q} \frac{[\bm{k^\prime}\times \bm{q}]^4}{q^4 \Pi^{(0)}_q}
 \mathcal{G}_{\bm{k^\prime} -\bm{q}}\mathcal{G}_{\bm{k} -\bm{q}} \right] \delta \Sigma_{\bm{k^\prime}} .
 \label{dPolar}
 \end{equation}
It is worthwhile to mention that the linear operator $\hat \alpha$ conserves the angular momentum, as follows from the rotational invariance of Eq.~\eqref{dPolar}. Therefore, it is convenient to split $\hat \alpha$ into the zeroth and second harmonics:
\begin{equation}
\hat \alpha k_x^2 = \frac{1}{2}
\hat \alpha_+ k^2 + \frac{k_x^2-k_y^2}{2k^2} \hat \alpha_- k^2 .
\end{equation}
The formal solution of Eq. \eqref{polar} can be then written as
\begin{equation}
\delta \Sigma_{\bm{k}} =- \frac{\delta \sigma}{2} \left(  \frac{1}{1+\hat \alpha_+}  + \frac{k_x^2-k_y^2}{k^2} \frac{1}{1+\hat \alpha_-} \right) k^2 .
\label{eq:deltaSigma}
  \end{equation}
Although Eq. \eqref{eq:deltaSigma} yields a formal solution for $\delta \Sigma_{\bm{k}}$, it is not justified to keep $\hat \alpha_\pm$ beyond the lowest order: not all the terms of the order $1/d_c^2$ can be generated from the diagram in Fig.~\ref{Fig-SE}. After a straightforward calculation, we obtain
\begin{equation}
\nu \approx - \frac13 +\frac 49 \bigl \langle \hat \alpha_+ - \hat \alpha_- \bigr \rangle_k ,
\label{eq:dPR:Alt}
\end{equation}
where
\begin{equation}
 \langle \hat \alpha_\pm \rangle_k = \frac{\int d^2\bm{k}\, k^2 \mathcal{G}_{\bm{k}}^2 \hat \alpha_\pm k^2}{\int d^2 \bm{k} \,k^4\mathcal{G}^2_{\bm  k} } .
 \label{<A>}
 \end{equation}
Expressing the difference $\langle \hat \alpha_+ - \hat \alpha_- \rangle_k$ in the rotationally invariant way, we obtain from Eq. \eqref{eq:dPR:Alt} exactly the same expression as in Eqs. \eqref{eq:gen:corr} and \eqref{eq:diff:eq1}.

%%%%%%%%%%%%%%%%%%%%%%%%%%
%%%%%%%%%%%%%%%%%%%%%%%%%%%

\subsection{Evaluation of the vertex corrections}

As we shall see below, all the integrals determining the $1/d_c$ correction to the differential Poisson's ratio are dominated by the momenta of the order of $q_\sigma$. Since the dependence of the bending rigidity on $q$ is controlled by $\eta\simeq 2/d_c$, we can neglect this dependence in the calculation of the  correction \eqref{eq:gen:corr}. Therefore, in what follows, we approximate the propagator of the flexural phonons by Eq. \eqref{eq:GG} with the bare bending rigidity.
Then, we find
\begin{equation}
\Pi^{(0)}_q = \frac{d_c T}{16\pi \varkappa^2 q^2}
\mathcal{P}\left(\frac{q\sqrt{\varkappa}}{\sqrt\sigma}\right ),
\end{equation}
where the dimensionless function $\mathcal{P}(Q)$ is given as
\begin{equation}
\mathcal{P}(Q) = \frac{8}{3} Q^2 \int\frac{d^2\bm{K}}{(2\pi)^2}
\frac{[\bm{K}\times \bm{Q}]^4}{Q^4}
\frac{1}{K^2(K^2+1)} \frac{1}{|\bm{Q}-\bm{K}|^2(|\bm{Q}-\bm{K}|^2+1)}  .
\end{equation}
The function $\mathcal{P}(Q)$ can be evaluated exactly with the help of the following set of transformations:
\begin{gather}
\mathcal{P}(Q) = \frac{8}{3} Q^2 \int\limits_0^\infty dt_1 dt_2 \left [ 1- e^{-t_1}\right ]\left [1-e^{-t_2}\right ] \int \frac{d^2\bm{K}}{(2\pi)^2}
\frac{[\bm{K}\times \bm{Q}]^4}{Q^4} e^{-t_1 K^2 -t_2 |\bm{K}-\bm{Q}|^2} =
Q^2  \int\limits_0^\infty \frac{dt_1 dt_2}{(t_1+t_2)^3}   \notag \\
\times \left [ 1- e^{-t_1}\right ]\left [1-e^{-t_2}\right ] e^{-Q^2 \frac{t_1 t_2}{t_1+t_2}} = \frac{Q^2}{4} \int\limits_{-\infty}^\infty \frac{dz}{\cosh^4z} \int_0^\infty \frac{d\tau}{\tau^2} e^{-Q^2\tau/2}
\prod\limits_{\sigma=\pm}
\left [ 1-e^{-\tau e^{\sigma z} \cosh z}\right ]
% \left [ 1-e^{-\tau e^{-z} \cosh z}\right ]
\notag \\
=
\frac{Q^4}{8} \int\limits_{-\infty}^\infty \frac{dz}{\cosh^4z} \Biggl\{ \left (1+\frac{4\cosh^2 z}{Q^2}\right )\ln \left (1+\frac{4\cosh^2 z}{Q^2}\right ) - 2  \left (1+\frac{2e^z\cosh z}{Q^2}\right )\ln  \left (1+\frac{2e^z\cosh z}{Q^2}\right )\Biggr\} \notag \\
= \frac{1}{3} \Biggl \{1+Q^4 \ln Q -(1+Q^2)^3\frac{\ln(1+Q^2)}{Q^2}  +Q(4+Q^2)^{3/2} \ln \frac{\sqrt{4+Q^2}+Q}{2}
\Biggr\} .
\label{eq:F}
\end{gather}
Here we used the parameterization $t_{1,2}=\tau e^{\pm z}\cosh z$. We note that the function $\mathcal{P}(Q)$ has the following asymptotic behavior:
\begin{equation}
\mathcal{P}(Q) = \begin{cases}
\displaystyle \frac{Q^2}{2}-\frac{Q^4}{6}(1-2\ln Q), & \quad Q\ll 1\\
\displaystyle  1-\frac{1}{2Q^2} (1+4\ln Q), & \quad Q\gg 1 .
\end{cases}
\end{equation}
In particular, we find that $\Pi_{xy}^{(0)} = d_c T/(32 \pi \varkappa\sigma)$.

Now we compute the contribution $I^{(a)}$ in Eq. \eqref{eq:gen:corr} from the diagram in Fig. \ref{fig1}a. This contribution can be written as
\begin{gather}
I^{(a)} = - \frac{2 d_c}{27\Pi^{(0)}_{xy}}
\int \frac{d^2 \bm{k}d^2 \bm{q}}{(2\pi)^4 T^2} \mathcal{G}_k^2\mathcal{G}_{\bm{k}-\bm{q}}^2 \frac{[\bm{k}\times \bm{q}]^6}{q^4}\frac{1}{\Pi^{(0)}_{q}} = -\frac{(32\pi)^2}{27 d_c}
\int \frac{d^2 \bm{Q}}{(2\pi)^2} \frac{Q^2}{\mathcal{P}(Q)} \mathcal{Y}_1(Q) ,
\end{gather}
where the function $\mathcal{Y}_1(Q)$ is given by
\begin{gather}
\mathcal{Y}_1(Q)=
\int\limits_0^\infty dt_1 dt_2 \left (\prod_{j=1,2} \Bigl [ t_j-2+(2+t_j)e^{-t_j}\Bigr ]\right ) \int \frac{d^2 \bm{K}}{(2\pi)^2} \frac{[\bm{K}\times \bm{Q}]^6}{Q^4} e^{-t_1 K^2 -t_2 |\bm{K}-\bm{Q}|^2} \notag\\
= \frac{15}{32\pi} \int\limits_0^\infty \frac{dt_1 dt_2}{(t_1+t_2)^4}  \left (\prod_{j=1,2} \Bigl [ t_j-2+(2+t_j)e^{-t_j}\Bigr ]\right ) e^{-Q^2\frac{t_1 t_2}{t_1+t_2}} .
\end{gather}
Performing the transformation $t_{1,2}=\tau e^{\pm z}\cosh z$ and integrating over $\tau$, we find
\begin{gather}
\mathcal{Y}_1(Q)= \frac{15 Q^2}{256\pi}
\int\limits_{-\infty}^\infty \frac{dz}{\cosh^6 z} \Biggl\{ \Bigl ( (1+2Q^2) \cosh^2 z+\frac{Q^4}{2}\Bigr ) \ln \frac{Q^4 + 4(1+Q^2) \cosh^2 z}{Q^4 + 4 Q^2\cosh^2 z}-2 \cosh^2 z\Biggr \}  \notag \\
= -\frac{1}{32\pi Q^2}\Biggl \{Q^4(5+10Q^2+2Q^4) \ln Q+ (1+Q^2)^2 \Bigl [ 2+ \bigl (2-9Q^2+6Q^4+2Q^6\bigr )\frac{\ln(1+Q^2)}{Q^2}\Bigr ]
\notag \\
+Q\sqrt{4+Q^2} (-10-3Q^2+6Q^4+2Q^6) \ln \frac{\sqrt{4+Q^2}+Q}{2}\Biggr \} .
\label{eq:Y}
\end{gather}

The contribution $I^{(b+c)}$ in Eq. \eqref{eq:gen:corr} from the diagrams in Fig. \ref{fig1}b and Fig. \ref{fig1}c can be computed in a similar way. We rewrite $I^{(b+c)}$ as follows:
\begin{gather}
I^{(b+c)}= \frac{(32\pi)^3}{81 d_c} \int\limits_0^\infty dt_1dt^\prime_1dt_2dt^\prime_2
\left (\prod_{j=1,2} \Bigl [ t_j-2+(2+t_j)e^{-t_j}\Bigr ][1-e^{-t_j^\prime}] e^{-Q^2 \frac{t_j t_j^\prime}{t_j + t_j^\prime}} \right )
\int \frac{d^2 \bm{Q}}{(2\pi)^2}\frac{Q^4}{\mathcal{P}^2(Q)}\notag  \\
\times
\int \frac{d^2 \bm{K_1}d^2 \bm{K_2}}{(2\pi)^4}
 [\bm{K_1}\times \bm{K_2}]^2
\left (\prod_{j=1,2}\frac{[\bm{K_j}\times \bm{Q}]^4}{Q^4} e^{-(t_j+t_j^\prime) (\bm{K_j} - \bm{Q} \frac{t_j^\prime}{t_j+t_j^\prime})^2}\right )
   .
\end{gather}
Then, integrating over $\bm{K_1}$ and $\bm{K_2}$, we get
\begin{gather}
I^{(b+c)} = \frac{2(32\pi)^3}{81 d_c}
\int \frac{d^2 \bm{Q}}{(2\pi)^2} \frac{Q^4}{\mathcal{P}^2(Q)}\mathcal{Y}_2(Q)
\Bigl [ \mathcal{Y}_2(Q)+\mathcal{Y}_3(Q)\Bigr ] ,
\end{gather}
where
\begin{gather}
\mathcal{Y}_2(Q)=
\frac{15}{32\pi} \int\limits_0^\infty \frac{dt_1 dt_1^\prime}{(t_1+t_1^\prime)^4}  \Bigl [ t_1-2+(2+t_1)e^{-t_1}\Bigr ]
\Bigr [1-e^{-t_1^\prime}\Bigr ]
 e^{-Q^2\frac{t_1 t_1^\prime}{t_1+t_1^\prime}} ,
\end{gather}
and
\begin{gather}
\mathcal{Y}_3(Q)=
\frac{3}{16\pi} \int\limits_0^\infty \frac{dt_1 dt_1^\prime}{(t_1+t_1^\prime)^4}  \Bigl [ t_1-2+(2+t_1)e^{-t_1}\Bigr ]
\Bigr [1-e^{-t_1^\prime}\Bigr ]
 \Bigl [-2+ Q^2 \frac{ t_1^{\prime 2}}{t_1+t_1^\prime}\Bigr ]
 e^{-Q^2\frac{t_1 t_1^\prime}{t_1+t_1^\prime}}
  .
\end{gather}
Using the parametrization $t_{1}=\tau e^{z}\cosh z$ and $t_{1}^\prime=\tau e^{-z}\cosh z$, and integrating over $\tau$, we obtain
\begin{gather}
\mathcal{Y}_2(Q) = \frac{15}{256\pi} \int_{-\infty}^\infty \frac{dz}{\cosh^6 z} \Biggl\{
- \cosh^2 z - \frac{Q^2}{4} (Q^2+1+e^{-2z})\ln \left (1+\frac{1+e^{-2z}}{Q^2}\right ) \notag \\ +
(Q^2+1+e^{-2z})\Bigl(\frac{Q^2}{4}+ \cosh^2 z\Bigr ) \Bigl [ \ln (Q^2+4 \cosh^2 z) - \ln(Q^2+1+e^{-2z}) \Bigr ] \Biggr\}
 .
\end{gather}
Integrating over $z$, we arrive at
\begin{gather}
\mathcal{Y}_2(Q) = \frac{1}{256\pi} \Biggl \{ - \frac{2}{Q^4}\bigl (6+7Q^2+6Q^4\bigr) - \frac{4}{Q^6} \bigl (1+Q^2\bigr )^3\bigl (2Q^4+4Q^2-3\bigr)\ln\bigl (1+Q^2\bigr) \notag \\+ 4 Q^2\bigl(2Q^2+5\bigr )\ln Q
+\frac{4}{Q} \bigl (2Q^2+3\bigr )\bigl (4+Q^2\bigr )^{3/2} \ln \frac{\sqrt{4+Q^2}+Q}{2}\Biggr\} .
\label{eq:A}
\end{gather}

The function $\mathcal{Y}_3(Q)$ can be conveniently expressed as
\begin{gather}
\mathcal{Y}_3(Q)=-\frac{4}{5}\mathcal{Y}_2(Q)+\tilde{\mathcal{Y}}_3(Q) ,
\end{gather}
where the function $\tilde{\mathcal{Y}}_3(Q)$ after the integration over $\tau$ acquires the following form:
\begin{gather}
\tilde{\mathcal{Y}}_3(Q) = \frac{3 Q^2}{512\pi}  \int\limits_{-\infty}^\infty \frac{dz}{\cosh^6 z} e^{-2z} \Biggl\{
 (2Q^2+1+e^{2z})\ln \left (1+\frac{1+e^{2z}}{Q^2}\right ) -
 (2Q^2+1+e^{-2z}+4 \cosh^2 z) \notag \\
 \times  \Bigl [ \ln (Q^2+4 \cosh^2 z)
 - \ln(Q^2+1+e^{-2z}) \Bigr ] \Biggr\} .
\end{gather}
Integration over $z$ yields
\begin{gather}
\tilde{\mathcal{Y}}_3(Q) = \frac{1}{160\pi} \Biggl\{ -Q^2(5+6Q^2) \ln Q + \frac{(1+Q^2)^2(6Q^6+8Q^4+8Q^2-9)}{Q^6}\ln(1+Q^2)
\notag \\
+
\frac{18+11Q^2+18Q^4}{2Q^4}
 - \frac{\sqrt{4+Q^2}}{Q}(26+23Q^2+6Q^4)\ln \frac{\sqrt{4+Q^2}+Q}{2}
\Biggr \} .
\end{gather}
Then, we obtain the following expression
\begin{gather}
\mathcal{Y}_3(Q) = -\frac{Q^{-4}}{128\pi} \Biggl \{-3(2+Q^2+2Q^4)+
2 Q^6(1+2Q^2) \ln Q -2  (-3+3Q^2+2Q^4+2Q^6)\notag \\
\times \frac{(1+Q^2)^2}{Q^2}\ \ln(1+Q^2)
+2 Q^3 (8+7Q^2+2Q^4)\sqrt{4+Q^2} \ln \frac{\sqrt{4+Q^2}+Q}{2}
\Biggr \} .
\label{eq:AB}
\end{gather}

\subsection{Final result for the differential Poisson's ratio}

\begin{figure}
\centerline{\includegraphics[width=0.6\textwidth]{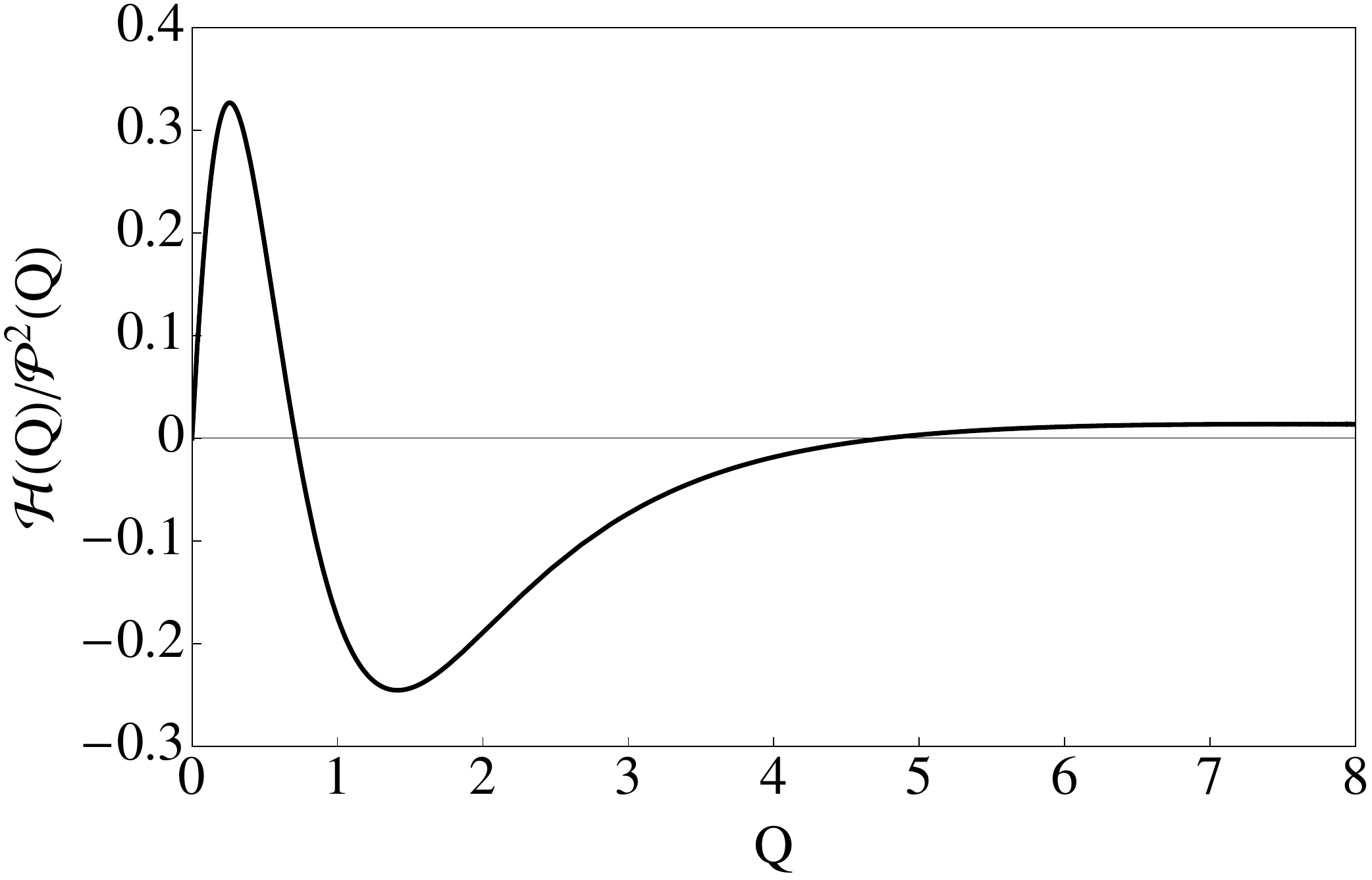}}
\caption{The plot of the function  $\mathcal{H}(Q)/\mathcal{P}^2(Q)$ (see text).}
\label{fig2}
\end{figure}

Combining together the results for the contributions $I^{(a)}$ and $I^{(b+c)}$, we express the difference of the polarization operators responsible to the $1/d_c$ correction to $\nu$ through a single integral:
\begin{gather}
\frac{3\Pi_{xy}-\Pi_{xx}}{9 \Pi^{(0)}_{xy}} = \frac{16}{81 d_c} \int\limits_0^\infty \frac{dQ \ \mathcal{H}(Q)}{\mathcal{P}^2(Q)},
\label{eq:c2-final}
\end{gather}
where
\begin{equation}
\mathcal{H}(Q) = 96 \pi Q^3 \Bigl \{ -\mathcal{Y}_1(Q)\mathcal{P}(Q)+\frac{64\pi}{3} Q^2 \mathcal{Y}_2(Q)[\mathcal{Y}_2(Q)+\mathcal{Y}_3(Q)]\Bigr \} .
\end{equation}
Using Eqs. \eqref{eq:F}, \eqref{eq:Y}, \eqref{eq:A}, and \eqref{eq:AB}, we obtain the following lengthy explicit expression for $\mathcal{H}(Q)$:
\begin{align}
\mathcal{H}(Q) = & -\frac{1}{8 Q^3} \Biggl \{
-4 Q^{12} (5 + 8 Q^2) \ln^2 Q +
    4 (1 + Q^2)^6 (9 - 30 Q^2 + 26 Q^4)\frac{\ln^2(1 + Q^2)}{Q^4}
    \notag \\
    &-
    4 (1 + Q^2)^2 \ \frac{\ln(1 + Q^2)}{Q^2}
     \Bigl [ 18 + 3 Q^2 - 26 Q^4 - 34 Q^6 - 58 Q^8 - 20 Q^{10}
     \notag \\
     &  +
       2 Q^3 \sqrt{4 + Q^2} \bigl (-30 - 3 Q^2 + 97 Q^4 + 38 Q^6 + 4 Q^8\bigr )
   \ln \frac{\sqrt{4 + Q^2}+Q}{2} \Bigr ]
   \notag \\
   & + \Bigl (36 + 60 Q^2 + 77 Q^4 + 28 Q^6 + 20 Q^8 -
      4 Q^3 \sqrt{4 + Q^2} \bigl (60 + 96 Q^2 + 143 Q^4
      \notag \\
     & + 100 Q^6 + 20 Q^8\bigr ) \ln \frac{\sqrt{4 + Q^2}+Q}{2}
       +   4 Q^6 (4 + Q^2)^3 (11 + 8 Q^2) \ln^2 \frac{\sqrt{4 + Q^2}+Q}{2} \Bigr )
\notag \\
& +
    4 Q^4 \Bigl (2 (1 + Q^2)^2 (9 + 5 Q^2 - 6 Q^4 + 4 Q^6) \ln(1 + Q^2)
       -       Q^2 \bigl (18 + 37 Q^2
       \notag \\
&       + 56 Q^4 + 20 Q^6 -
          2 Q^3 \sqrt{4 + Q^2} (16 + Q^2) \ln \frac{\sqrt{4 + Q^2}+Q}{2}\bigr )\Bigr )\ln Q
          \Biggr \} .
\end{align}
The function $\mathcal{H}(Q)$ has the following asymptotic behavior:
\begin{equation}
\mathcal{H}(Q) = \begin{cases}
5 Q^5/8,\quad  & Q\ll 1 ,\\
\displaystyle\left (\frac{485}{72} - \frac{65}{3} \ln Q + 10 \ln^2 Q\right )/Q^3, \quad & Q\gg 1 .
\end{cases}
\end{equation}
The function $\mathcal{H}(Q)/\mathcal{P}^2(Q)$ is shown in Fig. \ref{fig2}. As one can see, it changes sign twice which leads to a partial compensation of the corrections from diagrams on Fig. \ref{fig1}a-c. Numerically evaluating the integral in Eq. \eqref{eq:c2-final} and substituting it into Eq.~\eqref{eq:nu-with-corr}, we find the result \eqref{eq:nu:main}.

%%%%%%%%%%%%%%%%%%%%%%%%
%%%%%%%%%%%%%%%%%%%%%%%%

%%%%%%%%%%%%%%%%%%

\section{Conclusions\label{s3}}

To summarize, we have computed the differential Poisson's ratio of a suspended two-dimensional crystalline membrane embedded into a space of large dimensionality $d \gg 1$. Our result \eqref{eq:nu:main} demonstrates that, for $\sigma_L \ll \sigma \ll \sigma_*$, the differential Poisson's ratio of a crystalline membrane is a universal but non-trivial function of $d_c$. This results invalidates a  common belief (based on results of the self-consistent screening approximation) that the Poisson's ratio is equal to $-1/3$ independently of $d_c$. 

In the physical case of a two-dimensional membrane ($d_c =1$), one may speculate that the differential Poisson's ratio is not too far from the value $-1/3$ since the correction of the order $1/d_c$ in Eq. \eqref{eq:nu:main} is numerically  small. Clearly, a comparison with computational results would be of great interest. Unfortunately, the existing numerical results the Poisson's ratio of two-dimensional membranes (including graphene) are, however, quite controversial. This may be partly related with a very delicate character of the problem, see a detailed analysis in Ref. \cite{our-PR-paper}. As has been mentioned in Sec.~\ref{sec:intro}, the Poisson ratio in the linear-response regime $\sigma \ll \sigma_L$ depends on boundary conditions. In order to get rid of such finite-size effects but still to be in the regime of universal elasticity, the stress should be in the intermediate range $\sigma_L \ll \sigma \ll \sigma_*$. To resolve well this regime in numerical simulations, sufficiently large systems should be considered. Furthermore, in this regime, a care should be exerted in order to distinguish between the differential and the absolute Poisson ratio \cite{our-PR-paper}. 

Finally, we mention that it would be interesting to extend our analytical result for the $1/d_c$-expansion of the  differential Poisson's ratio of a two-dimensional membrane in two directions. First, one can address in a similar way the absolute Poisson ratio. (In this case, the zeroth-order term corresponding to the limit $d_c= \infty$ is equal to $-1$, see Ref.~\cite{our-PR-paper}.)  Second, the case of a disordered membrane \cite{my-crump} is of interest. 

\section{Acknowledgements}

We are grateful to E. Kats, M. Katsnelson, I. Kolokolov, V. Lebedev, and J. Los  for useful discussions.  The work was funded in part by Deutsche Forschungsgemeinschaft, by the Alexander von Humboldt Foundation, and by Russian Science Foundation under the grant No. 14-42-00044.

\appendix

\section{Screening of the elastic modulus $\mu$ and $\lambda$ \label{App1}}

In this Appendix, we present technical details of the calculation of screening of elastic modulus. We start from rewriting the term in Eq. \eqref{eq:action:iv} which describes the interaction between flexural phonons in a symmetric form \cite{Doussal}:
\begin{gather}
\frac{1}{4} \int\left ( \prod_{j=1}^4 \frac{d^D \bm{k_j}}{(2\pi)^2}\right )
\delta \left ( \sum_{j=1}^4 \bm{k_j}\right )
R_{\alpha\beta,\gamma\delta}(\bm{k_1}+\bm{k_2})\bigl (\bm{h}_{\bm{k_1}} \bm{h}_{\bm{k_2}}\bigr )
\bigl( \bm{h}_{\bm{k_3}} \bm{h}_{\bm{k_4}}\bigr ) .
\end{gather}
Here we consider a membrane of dimensionality $D$. The interaction kernel reads
\begin{gather}
R_{\alpha\beta,\gamma\delta}(\bm{q}) =
\frac{N}{D-1} P_{\alpha\beta}P_{\gamma\delta} + \mu \left (\frac{P_{\alpha\gamma}P_{\beta\delta} +P_{\alpha\delta}P_{\beta\gamma} }{2} - \frac{P_{\alpha\beta}P_{\gamma\delta} }{D-1} \right ) ,
\label{app:R:eq}
\end{gather}
where $N=\mu(2\mu+D\lambda)/(2\mu+\lambda)$. The projection operator is given as
\begin{equation}
P_{\alpha\beta} = \delta_{\alpha\beta} - \frac{q_\alpha q_\beta}{q^2} .
\end{equation}
The screened interaction kernel obeys \cite{Doussal}:
\begin{gather}
\tilde{R}_{\alpha\beta,\gamma\delta}(\bm{q}) =
R_{\alpha\beta,\gamma\delta}(\bm{q}) - R_{\alpha\beta,\gamma^\prime\delta^\prime}(\bm{q})
\hat{\Pi}_{\gamma^\prime\delta^\prime,\alpha^\prime\beta^\prime}(\bm{q}) \tilde{R}_{\alpha^\prime\beta^\prime,\gamma\delta}(\bm{q}) .
\label{app:SCSA:eq}
\end{gather}
The polarization operator at finite momenta can be written as \cite{lower-cr-D2}
\begin{gather}
\hat{\Pi}_{\gamma^\prime\delta^\prime,\alpha^\prime\beta^\prime}(\bm{q}) =
\Pi_{xy}(q) \delta_{\gamma^\prime\delta^\prime} \delta_{\alpha^\prime\beta^\prime}
+ \frac{1}{D} \bigl (\Pi_{xx}(q) -\Pi_{xy}(q)\bigr )
\Bigl ( \delta_{\gamma^\prime\alpha^\prime} \delta_{\delta^\prime\beta^\prime}+
\delta_{\gamma^\prime\beta^\prime} \delta_{\delta^\prime\alpha^\prime}\Bigr )
+\Pi_1(q) \Bigl ( \delta_{\gamma^\prime\delta^\prime}q_{\alpha^\prime}q_{\beta^\prime}
\notag \\
+
 \delta_{\alpha^\prime\beta^\prime}q_{\gamma^\prime}q_{\delta^\prime}
 \Bigr )
+ \Pi_2(q) \Bigl ( \delta_{\gamma^\prime\beta^\prime} q_{\delta^\prime}q_{\alpha^\prime}+\delta_{\gamma^\prime\alpha^\prime} q_{\delta^\prime}q_{\beta^\prime} +
\delta_{\delta^\prime\alpha^\prime}
 q_{\gamma^\prime}q_{\beta^\prime} +
\delta_{\delta^\prime\beta^\prime}
 q_{\gamma^\prime}q_{\alpha^\prime}
\Bigr ) +\Pi_3(q) q_{\alpha^\prime}q_{\beta^\prime}  q_{\gamma^\prime} q_{\delta^\prime} .
\notag \\
\end{gather}
Because of the projection operators entering $R_{\alpha\beta,\gamma^\prime\delta^\prime}$, the components $\Pi_1(q)$, $\Pi_2(q)$, and $\Pi_3(q)$ of the polarization operator drop from Eq. \eqref{app:SCSA:eq}. This equation can be solved by $\tilde{R}_{\alpha\beta,\gamma\delta}$ which has exactly the same structure as ${R}_{\alpha\beta,\gamma\delta}$, Eq. \eqref{app:R:eq},  but with the screened coefficients $N^\prime$ and $\mu^\prime$ instead of $N$ and $\mu$, respectively:
\begin{equation}
\mu^\prime(q)= \frac{\mu}{1+\Bigl (\Pi_{xx}(q)-\Pi_{xy}(q)\Bigr )\mu}, \quad
N^\prime(q) = \frac{N}{1+ \Bigl (2\Pi_{xx}(q)+(D-2)(D+1)\Pi_{xy}(q) \Bigr )N/D} .
\end{equation}
Within the self-consistent screening approximation the following relation holds: $\Pi_{xx}(q) = (D+1)\Pi_{xy}(q) \equiv (D+1)\Pi_q^{(0)}$, and we reproduce the results of Ref. \cite{Doussal}.

For $D=2$, we can rewrite these equations in the following way:
\begin{equation}
\mu^\prime(q) =\frac{\mu}{1+ \Bigl(\Pi_{xx}(q)-\Pi_{xy}(q)\Bigr)\mu} , \qquad
B^\prime(q) = \frac{B}{1+\Bigl (\Pi_{xx}(q)+\Pi_{xy}(q)\Bigr) B} .
\label{app:eq:screened:elmod}
\end{equation}
The result \eqref{app:eq:screened:elmod} generalizes Eq. \eqref{eq:screened:elmod} to a finite momentum transfer.

\section{Ward identity \label{App2}}

In this Appendix we discuss the Ward identity for the elastic action and its consequences for 
small-momentum
behaviour of exact propagators of flexural phonons.  While  the main text focuses on the  high-temperature regime, here we discuss a more general case of  arbitrary temperatures. For the sake of simplicity, we consider the case $d=3$.

\subsection{Basic equations}

We start from the following imaginary-time Lagrangian written in terms of the $3$-dimensional vector $\bm{r}$:
\begin{gather}
\mathcal{L}[\bm{r}] =   \rho (\partial_\tau \bm{r})^2+ \frac{\varkappa}{2} (\triangle \bm{r})^2 +\frac{\mu}{4} \Bigl (\partial_\alpha \bm{r} \partial_\beta \bm{r}- \delta_{\alpha\beta}\Bigr )^2
+ \frac{\lambda}{8} \Bigl (\partial_\alpha \bm{r} \partial_\alpha \bm{r}- 2 \Bigr )^2  .
\label{eq:S:1}
\end{gather}
Here Greek indices correspond to the 2D coordinates $(x,y) \equiv \bm{x}$ parameterizing the membrane. We note that substituting $\bm{r} = \{\xi_x x+u_x,\xi_y y + u_y, h\}$ into Eq. \eqref{eq:S:1} yields the membrane action \eqref{eq:action:i}.

The Lagrangian \eqref{eq:S:1} is manifestly invariant under $O(3)$ rotations of the vector $\bm{r}$. These rotations can be parameterized as
\begin{equation}
r_j \to r_j + \varepsilon_a t^a_{jk} r_k ,
\label{eq:Rot:1}
\end{equation}
where $\varepsilon^a \to 0$ are some constants and $t_{jk}^a=\epsilon_{ajk}$ are generators of $O(3)$ group. In order to explore implications of this
symmetry, we shall follow a standard approach \cite{S1,S2}. Let us consider the functional $\Phi[\hat \Sigma]$ defined as follows
\begin{equation}
\exp \Bigl ( - \Phi[\hat \Sigma] \Bigr ) = \int D[\bm{r}] \exp\Biggl \{ - \int\limits_0^\beta d\tau \int d^2x\, \Bigl ( \mathcal{L}[\bm{r}]- \Sigma_{j\alpha} \partial_\alpha r_j \Bigr ) \Biggr \} .
\end{equation}
At this stage, $\Sigma_{j\alpha}$ are arbitrary functions of $x$ and $y$; as will become clear soon, they have a meaning of components of the stress tensor $\sigma_{j\alpha}$ \cite{lower-cr-D2,david2}. The average deformation
\begin{gather}
\partial_\alpha {R_j} = \langle \partial_\alpha r_j \rangle
\end{gather}
can be found as
\begin{gather}
 \partial_\alpha {R_j} = - \frac{\delta \Phi[\hat \Sigma]}{\delta \Sigma_{j \alpha}} .
\label{eq:av:1-0}
\end{gather}
Evidently, $R_j$ transform according to Eq. \eqref{eq:Rot:1} under rotation.

Let us now consider the Legendre transform of $\Phi[\hat \Sigma]$:
\begin{gather}
\mathcal{F}[\bm{R}] = \Phi[\hat \Sigma]  +   \int\limits_0^\beta d\tau \int d^2x\, \Sigma_{j \alpha} \partial_\alpha R_j.
\label{eq:LL:1}
\end{gather}
Here $\Sigma_{j \alpha}$ should be found from the solution of Eq. \eqref{eq:av:1-0}. There is also the reciprocal relation between  $\Sigma_{\alpha j}$ and  $\partial_\alpha {R_j}$:
\begin{gather}
\Sigma_{j \alpha} = \frac{\delta \mathcal{F}[\bm{R}]}{\delta \partial_\alpha R_j} .
\label{eq:LL:1-2}
\end{gather}
We note that $\mathcal{F}[\bm{R}]$ coincides with the free energy evaluated under the constraint $\langle \partial_\alpha r_j \rangle = \partial_\alpha R_j$, where $\bm{R}$ is a given function of $x$ and $y$.

Now let us introduce the two-point correlation function ${\mathcal{S}_{jk}^{\alpha\beta}(\bm{q},\omega)}$ as the second variation of the functional $\mathcal{F}[{\bm{R}}]$:
\begin{equation}
{\mathcal{S}_{jk}^{\alpha\beta}(\bm{x}\tau,\bm{x^\prime}\tau^\prime)} =
 \frac{\delta^2 \mathcal{F}[{\bm{R}}]}{\delta {\partial_\alpha R_j}(\bm{x}\tau) \delta {\partial_\beta R_k}(\bm{x^\prime}\tau^\prime)} .
 \label{eq:SJ:1}
\end{equation}
We note that the propagator of displacements,
\begin{equation}
\mathcal{G}_{jk}(\bm{x}\tau,\bm{x^\prime}\tau^\prime) = - \bigl\langle\mathcal{T}_\tau r_j(\bm{x}\tau) r_k(\bm{x^\prime}\tau^\prime)\bigr \rangle_\Sigma ,
\end{equation}
where $\langle\cdots\rangle$ is defined with respect to the Lagrangian $\mathcal{L}[\bm{r}]- \Sigma_{j\alpha} \partial_\alpha r_j$ and $\mathcal{T}_\tau$ denotes the ordering along the imaginary time contour, is related with the two-point correlation function ${\mathcal{S}_{jk}^{\alpha\beta}(\bm{q},\omega)}$ in the following way:
\begin{equation}
\mathcal{G}^{-1}_{jk}(\bm{x}\tau,\bm{x^\prime}\tau^\prime) =
\frac{\partial^2}{\partial {x_\alpha} \partial{x^\prime_\beta}} {\mathcal{S}_{jk}^{\alpha\beta}(\bm{x}\tau,\bm{x^\prime}\tau^\prime)} .
\label{eq:S:Ginv}
\end{equation}

The rotation symmetry \eqref{eq:Rot:1} implies that
\begin{equation}
\Phi[\hat \Sigma] = \Phi[\hat \Sigma^\prime] ,
\end{equation}
where $\Sigma^\prime_{j \alpha} = \Sigma_{j \alpha} - \varepsilon_a t^a_{jk} \Sigma_{k \alpha}$.
Expanding this equation to the lowest order in $\varepsilon^a$, we find the Ward identity:
\begin{equation}
0 =
\varepsilon^a t^a_{jk}  \int\limits_0^\beta d\tau \int d^2x \, \Sigma_{k \alpha} \frac{\delta \Phi[\hat \Sigma]}{\delta \Sigma_{j \alpha}}  =
- \varepsilon^a t^a_{jk}  \int\limits_0^\beta d\tau \int d^2x\, \partial_\alpha R_j \frac{\delta \mathcal{F}[\bm{R}]}{\delta \partial_\alpha  R_k}  .
\label{eq:WT:1}
\end{equation}
In order to use the Ward identity for analysis of the two-point correlation function, it it convenient to perform a variation of the last part of Eq. \eqref{eq:WT:1} with respect to $\partial_\gamma R_l(\bm{x}^\prime,\tau^\prime)$. This yields
\begin{equation}
\varepsilon^a t^a_{lk} \Sigma_{k\gamma}(\bm{x}^\prime,\tau^\prime) +
\varepsilon^a t^a_{jk}  \int\limits_0^\beta d\tau \int d^2x\,  \partial_\alpha  R_j(\bm{x},\tau)
{\mathcal{S}_{kl}^{\alpha\gamma}(\bm{x}\tau,\bm{x^\prime}\tau^\prime)}
  = 0 .
\label{eq:WT:2}
\end{equation}

\subsection{The propagator of flexural phonons}

With the choice $\bm{\varepsilon}=\{\varepsilon, 0, 0\}$,  Eq. \eqref{eq:WT:2}  reduces to
\begin{equation}
t^x_{zy} \Sigma_{y\gamma}(\bm{x}^\prime,\tau^\prime)  + t^x_{yz}
 \int\limits_0^\beta d\tau \int d^2x\,  \partial_\alpha  {R_y(\bm{x},\tau)} \,\,
 {\mathcal{S}_{zz}^{\alpha\gamma}(\bm{x}\tau,\bm{x}^\prime\tau^\prime)}
  = 0 .
 \label{eq:WT:4}
\end{equation}
Now we consider the function $\bm{R}(\bm{x},\tau)$ which has the following form:
\begin{equation}
{\bm{R}}(\bm{x},\tau) = \bm{R}^{(\xi)}=\{{\xi}_x x, {\xi}_y y, 0\} ,
\label{eq:av:RR}
\end{equation}
where  $\xi_x$ and $\xi_y$ are arbitrary constants.
%In general, the choice \eqref{eq:av:RR} implies non-zero value of $\Sigma_{xx}$ and $\Sigma_{yy}$ in virtue of Eq. \eqref{eq:LL:1-2}. The following remarks are in order here.
The functional $\mathcal{F}[\bm{R}^{(\xi)}]$ corresponds to the free energy evaluated  under the constraint $\langle \partial_\alpha r_j \rangle =\xi_\alpha \delta_{\alpha j}$, where the average is taken with respect to the Lagrangian $\mathcal{L}[\bm{r}]$. This is exactly the action $S$ (see Eq. \eqref{eq:action:i}) discussed in the main text. Using Eq. \eqref{eq:WT:4}, we find
\begin{gather}
{\xi}_y \lim\limits_{\omega,q\to 0}{\mathcal{S}_{zz}^{yy}(\bm{q},\omega)} = \Sigma_{yy}=\frac{\partial {f}}{\partial {\xi}_y}, \qquad
{\xi}_y \lim\limits_{\omega,q\to 0}{\mathcal{S}_{zz}^{yx}(\bm{q},\omega)} =0 ,\notag \\
{\xi}_x \lim\limits_{\omega,q\to 0}{\mathcal{S}_{zz}^{xx}(\bm{q},\omega)} = \Sigma_{xx}=\frac{\partial {f}}{\partial{\xi}_x}, \qquad
{\xi}_x \lim\limits_{\omega,q\to 0}{\mathcal{S}_{zz}^{xy}(\bm{q},\omega)} =0 .
\end{gather}
We recall that the physical stress is defined by Eq. \eqref{eq:EOS}. Therefore, we obtain
\begin{equation}
\lim\limits_{\omega,q\to 0}\mathcal{S}_{zz}^{\alpha\gamma}(\bm{q},\omega) = \sigma_{\alpha}\delta_{\alpha\gamma} .
\end{equation}
By virtue of Eq. \eqref{eq:S:Ginv}, this implies that the inverse propagator of the flexural phonons for $q\to 0$ has the following exact form:
\begin{equation}
\lim\limits_{\omega\to 0}\mathcal{G}^{-1}_{zz}(\bm{q},\omega) = \sigma_{x} q_x^2+\sigma_{y} q_y^2 + \dots
\label{eq:WI:FP:final}
\end{equation}
We note that Eq. \eqref{eq:WI:FP:final} extends the statement of Refs. \cite{lower-cr-D2,david2} to the case of $\sigma_{x}\neq \sigma_{y}$.

\vspace{1cm}

\end{document}